\documentstyle[aps,preprint]{revtex}
\begin{document}
\draft
\title{Diffusion with critically correlated traps and the slow
relaxation of longest wavelength mode}
\author{Sonali Mukherjee and Hisao Nakanishi}
\address{Department of Physics, Purdue University, W. Lafayette, IN 47907}
\date{\today}
\maketitle
\begin{abstract}
We study diffusion on a substrate with permanent traps distributed
with critical positional correlation, modeled by their placement
on the perimeters of a critical percolation cluster.  We perform
a numerical analysis of the vibrational density of states
and the largest eigenvalue of the equivalent scalar elasticity problem
using the method of Arnoldi and Saad.  We show that the critical trap
correlation increases the exponent appearing in the stretched
exponential behavior of the low frequency density of states
by approximately a factor of two as compared to the case of
no correlations.  A finite size scaling hypothesis of the largest eigenvalue
is proposed and its relation to the density of states is given.
The numerical analysis of this scaling postulate leads to
the estimation of the stretch exponent in good agreement with
the density of states result.
\end{abstract}
\pacs{05.40.+j, 05.50.+q, 64.60.Fr}

\newpage
\section{Introduction}
\label{sec:Intro}
Understanding the behavior of a particle diffusing  in the presence
of traps  is important as it captures
the essence of many physical processes, including
diffusion controlled reactions where one of the reactants is immobile,
trapping of excitons, etc. \cite{kayser}
Mapping of the diffusion problem with traps
to the scalar elasticity problem also affords an insight into lattice
vibration of systems such as binary alloys with
contrasting elastic constants. From a theoretical point of view,
this problem is analogous to the ideal chain in inhomogeneous media.
The ideal chain problem is interesting as it is the Gaussian limit of the
self-avoiding walks and yet possesses a universality class of its own
in inhomogeneous media, distinct from the usual random walk.

These considerations resulted in many analytical and numerical attempts
to gain understanding of the problem \cite {danny}.
Typical quantities used to characterize diffusion in the
presence of traps include $ P_0 (t) $,
the probability that the diffusing particle
returns to the starting point after time $t$.
$ P_0 (t) $ is related to the number of distinct sites visited by the
diffusing particle and is the Laplace transform of the vibrational
density of states  of the corresponding scalar
elasticity problem, which can often be measured
experimentally by methods such as Raman and neutron scattering
\cite{aerogel}.

For the case of diffusion in the presence of traps distributed randomly
with no correlations,
it has been proved rigorously by Donsker and Varadhan \cite {donsker}
that the decay of $ P_0 (t) $ with time is slower than exponential,
indicating that, in the long time limit, the
properties of the diffusing particle is dominated by the
presence of large trap free regions which have a finite (though small)
probability of occurence. This is because, if the traps were uniformly
distributed, the diffusing particle would be trapped at a constant rate,
which would result in an exponential decay of $P_0 (t)$. It also means
that the {\em quenched} disorder average must be carried out since
an annealed averaging would smear out the trap positions.

In this paper we ask the question of whether and how this
interesting behavior of the diffusing particle with traps
is modified when we introduce long range
correlations in their positions \cite {achille}.
In the equivalent scalar elasticity problem, the correlated traps
map to the clamping of sites with a correlated distribution.
In this sense, the present problem extends the so-called {\em fractino}
problem \cite{sapoval} where a fractal boundary of an otherwise
nonfractal object is clamped to the case where the bulk of the substrate
is itself a fractal. It is also an extension of the {\em fracton}
problem \cite{alex} as the traps (or clamped boundaries) are
introduced into the scalar elasticity of fractals.

Calculational difficulties have prevented this problem from receiving
its share of attention even though, often in real physical situations,
the positions of traps are correlated at least within limited length scales.
For the case of uncorrelated trap distribution,
a Poisson distribution is usually assumed, which simplifies
the theoretical calculation \cite {grassberger}; however
this assumption fails for correlated traps. As far as computational
calculations, techniques based on exact enumeration are highly
computer time and memory intensive because
of the high sensitivity of the behavior of the diffusing particle
to the actual positions of the traps in the sample,
leading to large fluctuations
in the measured quantity from sample to sample.  This translates to
requiring an ensemble average to be taken over a large number of disorder
configurations for meaningful results. Moreover the onset of the
asymptotic regime is very slow in this type of problem,
which adds to the computational difficulties \cite {havlin}.

In this paper we establish  that $P_0 (t)$ has a stretched
exponential form {\em qualitatively} similar to the case of
the Poissonian trap distribution
but with a substantially different stretch exponent. We show this behavior
from the analysis of the density of eigenvalues of the transition
probability matrix ${\bf W}$ which describes the diffusion process.
We also study the finite size scaling of the largest eigenvalue
of ${\bf W}$, which serves as an extremely powerful tool to extract the
stretch exponent characterizing the long time behavior of $P_0 (t)$.
A preliminary account of this work was presented at the Hayashibara
Forum (1995) and appears in its proceedings. \cite{suzuki}

\section{Ideal chain in critically correlated disorder}
We choose percolation cluster \cite{perc} formed at critical
probability of occupation $ p_c $ as the substrate for diffusion.
The sites on the external perimeter (hull) \cite{hull} as well as on the
internal perimeter of the percolation cluster are made absorbing
and once the diffusing particle reaches the perimeter sites it
is trapped permanently. The hull sites and the sites constituting
the internal perimeter form fractals at $ p_c $ \cite{fractal}.
Thus placing the traps along the perimeters induces spatial correlations
among the traps because of the long range correlations among the sites
of a fractal.  While we do not pursue the distinction of external
versus internal perimeters in this work, there {\em are} interesting effects
when different boundary conditions are applied to them, at least
in two dimensions \cite{sonali2}.

The time evolution of the probability $P_i (t)$ that the diffusing particle
is at site $i$ at time $t$ is Markovian and the
process can be described in the continious time limit by the following
master equation
\begin{equation}
\label {eq:markov}
(\partial /{\partial t}) P_{i}(t) = \sum_{j} w_{ij}P_{j}(t) - P_{i}(t),
\end{equation}
where $w_{ij}$ is the hopping rate
from site $j$ to $i$. In this problem we consider only nearest neighbour
hopping and so $w_{ij}$ is non-zero only for the nearest neighbour pairs
$i,j$. We have chosen $w_{ij}$
to be $1/z$ for all occupied nearest neighbours $j$,
where $z$ is the full coordination number of the underlying lattice;
otherwise  $w_{ij}$ is set equal to zero.
What this amounts to is that
once the particle hops to a trap site there is no
further time evolution of that particular random walk and only those
walks which have escaped getting trapped in the perimeter sites
evolve further. Thus the traps act as permanent particle absorbers.

Processes whose time evolution is governed by Eq.\ (\ref{eq:markov})
can be cast into an eigenvalue problem
of the transition probability matrix ${\bf W}$.
The matrix elements $w_{ij}$ of ${\bf W}$ control the dynamics of the random
walk and the locations of the nonzero
elements of ${\bf W}$ have the information about the structure of the
underlying
fractal substrate responsible for the correlations
among the trap sites on its perimeters.

The density of normal modes of ${\bf W}$
is related to the return to the starting point probability of the diffusing
particle by the Laplace transform \cite {sonali}.
The probability distribution of the number of walks $C_0 (t)$ which return to
their starting point after time $t$, denoted as $P(C_0 (t);t)$,
was studied in \cite {hisao} and found to be a truncated
log-normal distribution. Thus
\begin{equation}
P( C_0 (t);t) \approx \frac{1}{C_0 (t) \sqrt{2\pi{\sigma_t}^2}}
{\exp}(\frac{-{(\ln C_0 (t) -\lambda_t)}^2}{2{\sigma_t}^2}) .
\end{equation}
In \cite {hisao} it was further found from the first moment of this
distribution that
\begin{equation}
\label{eq:rtsp}
\ln P_0 (t) \sim -t^{2(1-{\chi}_0)}
\end{equation}
in the asymptotic long time limit,
where $P_0 (t)$ is obtained as $\overline{C_0 (t) /z^t}$, the bar above the
quantity indicating the quenched disorder average.
The exponent ${\chi}_0$ is the same one as that which
appears in the long time behavior of the width
of the log-normal distribution, ${\sigma_t}^2$,
\begin{equation}
\label{eq:chi}
{\sigma_t}^2 \sim \beta{t^{2\chi_0}} .
\end{equation}

Since $P_0 (t)$ has a stretched exponential behavior according to
Eq.\  (\ref{eq:rtsp}),
we expect the density of normal modes $\rho$ (which is the inverse Laplace
transform of $P_0 (t)$) to also have a stretched exponential form
\cite{hisao},
\begin{equation}
\ln \rho(\epsilon) \sim -\epsilon^{-d_0/2}
\end{equation}
in the limit of small $\epsilon \equiv |\ln \lambda|$,
where $\lambda$ denotes the eigenvalues of ${\bf W}$,
and the exponents $\chi_0$ and $d_0$ are related to each other
by $d_0 = 4(1-\chi_0)/(2\chi_0 -1)$.
The results from \cite{hisao} gave support for the presence of
such a behavior although their numerical estimates of $d_0$
need to be improved as they did not take into account
the proper normalization of $\rho$ as well
as the substantial non-asymptotic effects \cite{hisao2}.

We can also relate the behavior of $\rho (\epsilon )$ to
the finite size effects of the edge of the spectrum.
First, note that, for any finite substrate however large, the largest
eigenvalue of ${\bf W}$, which we denote by $\lambda_1$, must be less than one.
This is because the eigenvalue of one would indicate the existence
of a stationary state whereas there are always particles which are trapped,
leading to a {\em leakage} of probability for the diffusing particle in
contrast to the diffusion {\em without} traps.  The value of
$|\ln \lambda_1|^{-1}$ then corresponds to the slowest time scale of the
problem, and there is always a gap between $\lambda_1$ and one,
the latter being what the maximum eigenvalue would be
for diffusion without traps.
The interesting question is whether this gap is bounded from below
by a nonzero constant or it approaches zero as the substrate size increases.

Our argument for the behavior $\lambda_1 \rightarrow 1$ as
the substrate size $S \rightarrow \infty$ follows the observation
of \cite{donsker} for the case of uncorrelated trap distribution.
In that case, the return probability $P_0 (t)$ was shown to be
a stretched exponential which is slower than a pure exponential
(although not as slow as a power law which would be the case in
the absence of traps), and this behavior was attributed
to the dominance of the trap-free regions although they occur relatively
infrequently.  In our case of critically correlated trap distribution,
we also have a slow, stretched exponential for $P_0 (t)$ which we interpret
as a similar dominance of the nearly trap-free regions.
Although there are more traps in the critically correlated case,
there are also much greater fluctuations in their density.
Thus regions of relatively low concentrations of the traps are
likely to be present, probably with a hierachical size distribution.
Since the leakage of probability
is {\em nearly} zero in a {\em nearly} trap free region, we would expect
$\lambda_1$ to approach one as the size of the largest trap free
region grows indefinitely.

Another argument is as follows: if there were indeed a nonzero lower
bound for $\epsilon_1 \equiv |\ln \lambda_1|$, then it would
also give a bound for the slowest relaxation rate.  Thus $P_0 (t)$ would
have to decay at least exponentially in time corresponding to this rate.
Since $P_0 (t)$ in fact decays as a stretched exponential with the
stretch exponent less than one (\cite{hisao} as well as this work),
no such bound can exist for the rate; rather $\epsilon_1 \rightarrow 0$
as $S \rightarrow \infty$.

We further propose a scaling relation between $\epsilon_1$ and $S$ via
\begin{equation}
\label{eq:eg}
1/S \sim \int_{0}^{\epsilon_1} \rho(\epsilon)d(\epsilon) ,
\end{equation}
where a possible numerical prefactor has been neglected.
This relation follows from the assumption that, say, the largest ($\lambda_1$)
and second largest ($\lambda_2$) eigenvalues scale relative to the value one
in the same way when $S \rightarrow \infty$. That is, if we assume
\begin{eqnarray}
\label{eq:2eigen}
\int_{0}^{\epsilon_1} \rho(\epsilon)d(\epsilon) & \sim & C_1 f(S) ,\\
\label{eq:3eigen}
\int_{0}^{\epsilon_2} \rho(\epsilon)d(\epsilon) & \sim & C_2 f(S) ,
\end{eqnarray}
where $\epsilon_2 \equiv |\ln \lambda_2|$, then the difference must also
scale in the same way,
\begin{equation}
\label{eq:4eigen}
\int_{\epsilon_1}^{\epsilon_2} \rho(\epsilon)d(\epsilon)
\sim (C_2 -C_1) f(S) ,
\end{equation}
but this latter integral must scale as $1/S$ if the integral of
the density of normal modes $\rho(\epsilon)$ is normalized to unity.
This means $f(S) \sim 1/S$, thus Eq.\ (\ref{eq:eg}) follows.

The assumption of the same asymptotic behavior for the two largest
eigenvalues is plausible if we consider the difference between
$\lambda_1$ (or $\lambda_2$) and one to be the reflection of the finite size
of the largest (or the second largest) trap free region (respectively).
As long as the large trap free regions are geometrically similar
(as would be the case in a hierarchical distribution of such regions),
and as long as their sizes all go to infinity as $S \rightarrow \infty$,
it seems reasonable that the gaps between $\lambda_i$ ($i=1,2$) and one behave
in the same manner.

If we substitute a stretched exponential form of $\rho(\epsilon)$ in
Eq.\ (\ref{eq:eg})
we get an incomplete gamma function. On retaining only the
leading term of the incomplete gamma function and taking natural log
of both sides we get,
\begin{equation}
\label{eq:fs}
\ln S =  a\epsilon_1^{-x} -(x+1)\ln \epsilon_1 +b ,
\end{equation}
where a general stretched exponential form of $\rho(\epsilon)$
\begin{equation}
\label{eq:spec}
-\ln \rho(\epsilon) \sim a\epsilon^{-x} + c ,
\end{equation}
has been assumed and $b=c+\ln (ax)$.

While the above gives a particular scaling prediction for the case
of the critically correlated trap distribution, the argument leading
to Eq.\ (\ref{eq:eg}) applies more generally. For example,
a similar procedure should apply to the uncorrelated trap distribution
of \cite{donsker}. Also, for the trapless case (or the {\em ants}
\cite{sonali}), where the largest eigenvalue is actually one
(because there {\em is} a stationary state), a similar argument should work
with, say, the second ($\lambda_2$) and third largest ($\lambda_3$)
eigenvalues.  Of course, if there is no trap, a probability leakage, per se,
does not occur. However, each mode with a large $\lambda$ tends to be
associated with a {\em blob} of high connectivity region, to which
the same kind of argument can be applied.
Indeed, in the trapless case, Eq.\ (\ref{eq:eg}) together with the power law
density of states
\begin{equation}
\rho(\epsilon) \sim \epsilon^{d_s/2 -1} ,
\end{equation}
where $d_s$ is the spectral dimension of the substrate,
leads to a power law relation between $S$ and $\epsilon_2$,
\begin{equation}
\epsilon_2 \sim S^{-2/d_{s}} ,
\end{equation}
which is identical to the relation proposed on the basis of the finite size
scaling of the largest nontrivial normal mode and numerically verified in
\cite {sonali}.
Such finite size scaling relations provide a very
powerful technique to obtain the quantitative characterization of
the density of states as they reduce the computational effort drastically,
requiring only the information on the highest (or second highest) mode.

\section{Results of normal mode analysis}
\label{sec:results}
In this section we give numerical results for the exponent $x$
for the stretched exponential decay of the density of states
as discussed above. We extract this exponent directly from the
density of states (cf. Eq.\ (\ref{eq:spec}))
and also independently from the finite size scaling of the
largest eigenvalue of ${\bf W}$ (cf. Eq.\ (\ref{eq:fs})) in two and three
dimensions (square and simple cubic lattice, respectively).

In order to reduce the computational time and memory requirements,
we take advantage
of the fact that we are interested only in the asymptotic long time
relaxation of the system which is controlled by those normal modes
with large eigenvalues $\lambda$. Thus we use the Arnoldi-Saad algorithm
\cite{saad,sonali} to reduce the original ${\bf W}$ into a
smaller matrix which
contains the approximate information about the highest normal modes.

We analyze the density of states per site, $\rho(\epsilon)$.
It is obtained from the eigenvalues of ${\bf W}$ by
binning them linearly in the $\epsilon$ space. The number of eigenvalues
in each bin is divided by the bin width, the size of the cluster and also
by the number of clusters  over which the quenched disorder average is
performed. The number of
independent cluster realizations over which the disorder average was taken
is of the order $1000$ for cluster sizes $S=$  $8000$, $10000$, and
$50000$ and of the order $10000$ for $S=$ $5000$. We retained in the final
results only those bins which contained
eigenvalues contributed from every substrate realization to avoid partial
binning.

Since $\rho(\epsilon)$ has a stretched exponential decay we have plotted
$-\ln \rho(\epsilon)$ versus $\epsilon$ in Fig.~\ref{fig1} which is expected
to have the form $a{\epsilon}^{-x} + c$.
There is an excellent data collapse
for all the cluster sizes both in 2d and 3d. The solid curve
drawn through the data is obtained by fitting the data to an
expression of this form. The numerical estimates of
$x$ and the numerical values of $a$ and $c$ corresponding to the central values
of $x$ are tabulated in Table\ \ref{table1}. Note that the
value of $x$ cannot be accurately obtained simply from the slope of
the double log plot of $\ln \rho(\epsilon)$ with respect to $\epsilon$
as the value of the constant $c$ might be appreciable.

The size of the symbols in Fig.~\ref{fig1} is larger than the cluster to
cluster  statistical fluctuations.
The quoted error bars are obtained visually from changing
the effective estimate of $x$ for the nonlinear fit
until it no longer fits the data points.
The small scaling regime which is typical for these kinds
of problems (i.e., diffusion with traps) \cite{havlin} makes the precise
extraction of $x$ difficult. This accounts for the large error bars
in our results as compared to the case of diffusion in a percolation
cluster without absorbing sites \cite {sonali}. In the latter case
the density of states has a power law form and the scaling regime
increases appreciably with the size of the cluster, which resulted
in much smaller error bars for the extracted exponents even with the same
numerical technique.

In Fig.~\ref{fig2} we plot $\ln S $ with respect to
$\epsilon_1$ in 2d and 3d where $S$ is the size of the substrate.
The solid curves are obtained by fitting the data to an expression
$a{\epsilon_1}^{-x} -(x+1) \ln \epsilon_1 +b$, which is what we
expect from the finite size scaling analysis of
the largest eigenvalue. The size of the symbols are larger than
the cluster to cluster fluctuations. The number of clusters over which the
disorder average was performed is of the order 1000 for the larger
clusters, same as for the density of states, and for smaller clusters
of size $100$, $400$, $1000$ and $5000$ the average was taken over $10000$
clusters. The estimates of $x$, and the values of $a$, and $b$ corresponding
to the central values of $x$ which we obtain from the
nonlinear fit are tabulated in Table\ \ref{table2}.
The error bar for the value of $x$ is obtained in a similar way
to that for the density of states.

We note that the estimates of $x$ and $a$ obtained from the density
of states and from the finite size scaling of the largest eigenvalue,
given in Table\ \ref{table1} and \ref{table2}, respectively,
are in good agreement with each other. The values of $d_0$ thus
obtained, however, turn out to be about a factor of two larger
than the corresponding stretch exponents
for the uncorrelated distribution of traps \cite{donsker}.
The estimates of $x$ also differ from those of $d_0/2$ as
given by \cite{hisao} due, we believe, primarily to the failure
of \cite{hisao}to take proper account of $\epsilon$ being not quite
in the asymptotic region (thus proper normalization and prefactors
becoming important) (cf. \cite{hisao2}). However, the exponent $\chi_0$
(Eq.\ (\ref{eq:chi})) is relatively insensitive to $x$ ($=d_0/2$) and
the analysis of $P_0 (t)$ in \cite{hisao} is not affected by these
problems. We also believe that their main conclusions remain valid.

\section{Summary}
\label{sec:con}
In summary, we have studied the problem of diffusion on a substrate
with permanent traps which are distributed with critical correlation
in their positions. The critical correlation has been modeled by placing
the traps on the perimeters of critical percolation clusters, which
are obtained by Monte Carlo simulation.  The statistical behavior of
the diffusing particle in the time domain is then mapped to
the scalar elasticity problem with fixed, fractal boundaries, and
a numerical analysis of the vibrational density of states of the latter
problem is carried out using an approximate diagonalization algorithm
of Arnoldi and Saad \cite{saad}.
This problem may be considered a generalization
both of the {\em fracton} problem of Alexander and Orbach \cite{alex},
where there are no traps but the substrate is fractal,
and of the {\em fractino} problem of Sapoval et al \cite{sapoval}
where a fractal boundary is clamped (equivalent to traps) but
the substrate itself is not fractal.

We have shown that introducing critical spatial correlations among the
traps does not change the {\em qualitative} behavior of the density of states
as compared to the case of uncorrelated traps,
but that there are substantial effects of correlations {\em quantitatively}.
This strongly suggests that the diffusion process is dictated by the presence
of nearly trapless regions similarly to the case of uncorrelated
traps, but the long range correlations among the trap positions lead
to a decrease in the number and size of these regions nearly free of traps,
which induces a faster
decay in the density of states and consequently a smaller probability
of return to the starting point after time $t$. This is reflected in a
much larger exponent in the stretched exponential form of density of states
than the case of uncorrelated traps.

On the other hand, as compared to the case of the same {\em fractal}
substrate but with {\em no} traps, the density of states is
{\em qualitatively} different, since the latter problem
produces a power law density of states in the low energy limit
known as {\em fractons}, accumulating to an infinite density toward
the maximum eigenvalue of one \cite{sonali}.  In contrast,
with the perimeters acting as traps, the density of states becomes
a stretched exponential, rapidly falling to zero toward the maximum
eigenvalue.

\acknowledgments
We appreciate helpful discussions with A. Giacometti and B. Sapoval.
During this work, SM was supported by Purdue Research Foundation,
to which she expresses her gratitude.

\newpage
\begin{figure}
\caption{\label{fig1}
Natural log of the density of eigenvalues $\rho(\epsilon)$
against $\epsilon$ in $d=2$ and $d=3$. The data
collapse very well for different sizes of substrates.}
\end{figure}

\begin{figure}
\caption{\label{fig2}
Natural log of the size of substrate $\ln S$ against $\epsilon_1$
in $d=2$ and $d=3$.}
\end{figure}

\newpage
\begin{table}
\caption{Estimates of the exponent $x$ and the constants
$a$ and $c$ from $\rho(\epsilon)$ for the square lattice in
two dimensions and simple cubic lattice in three dimensions.
The data for $-\ln \rho(\epsilon)$ have been fitted to the form
$ a\epsilon^{-x} + c $.}
\label{table1}
\begin{tabular}{cccc}
$d$ & $x$ & $a$ & $c$ \\
\hline
 2 & $2\pm 0.5$ & 0.084 &  0.42  \\
 3 & $3.24^{+ 0.3}_{-0.7}$ & 0.64  & -0.077
\end{tabular}
\end{table}

\begin{table}
\caption{Estimates of the exponent $x$ and the constants
$a$ and $b=\ln (ax) + c$ from the finite size scaling of $\epsilon_1$,
for the square lattice in two dimensions and simple cubic lattice
in three dimensions. The data have been fitted to the form
$ a\epsilon^{-x} -(x+1)\ln\epsilon + b$.}
\label{table2}
\begin{tabular}{cccc}
$d$ & $x$ & $a$ & $b$ \\
\hline
 2 & $2.23^{+0.3}_{-0.7}$ & 0.072 & -1.06  \\
 3 & $3.26^{+0.3}_{-0.7}$ & 0.5   &  0.72
\end{tabular}
\end{table}

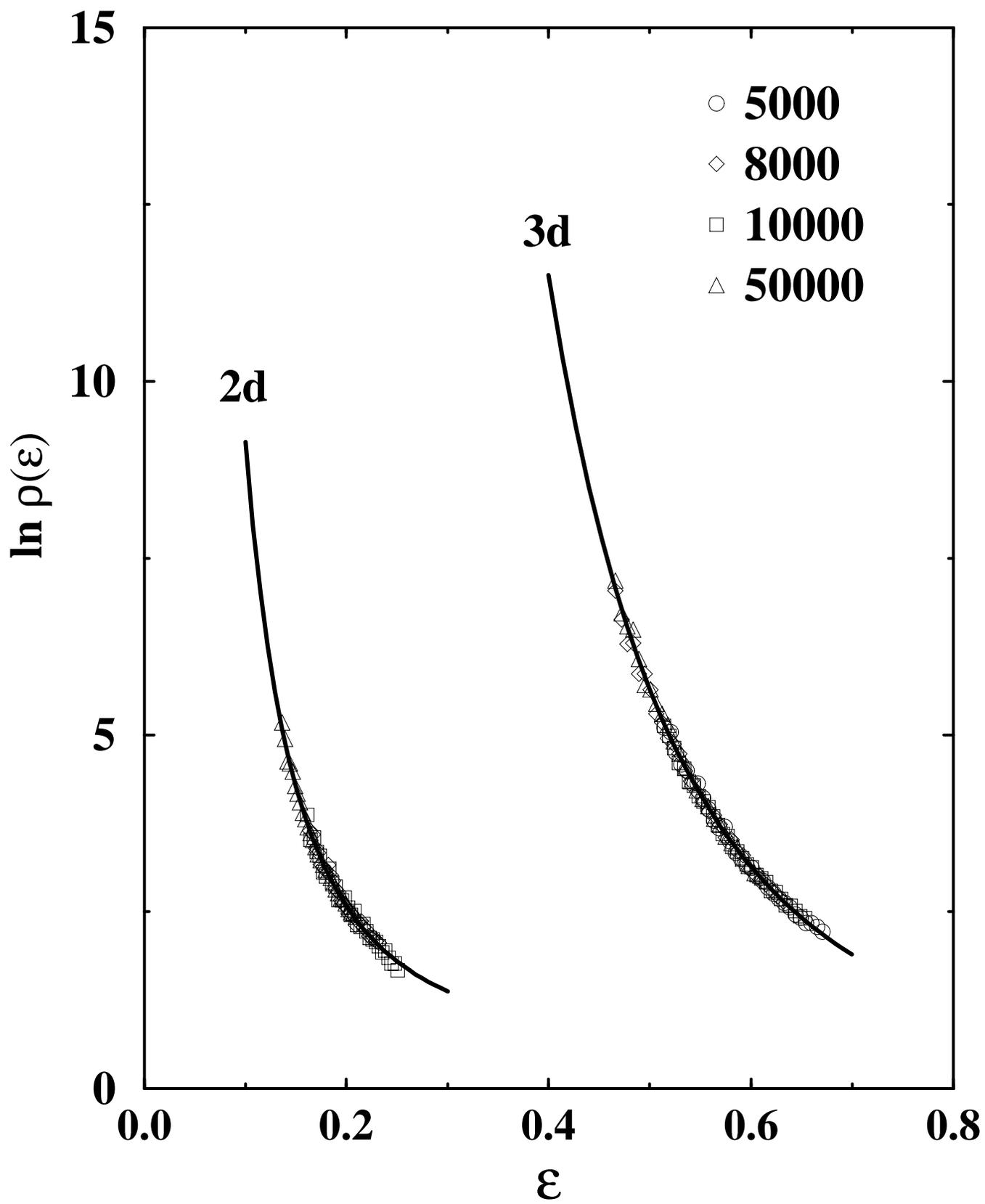
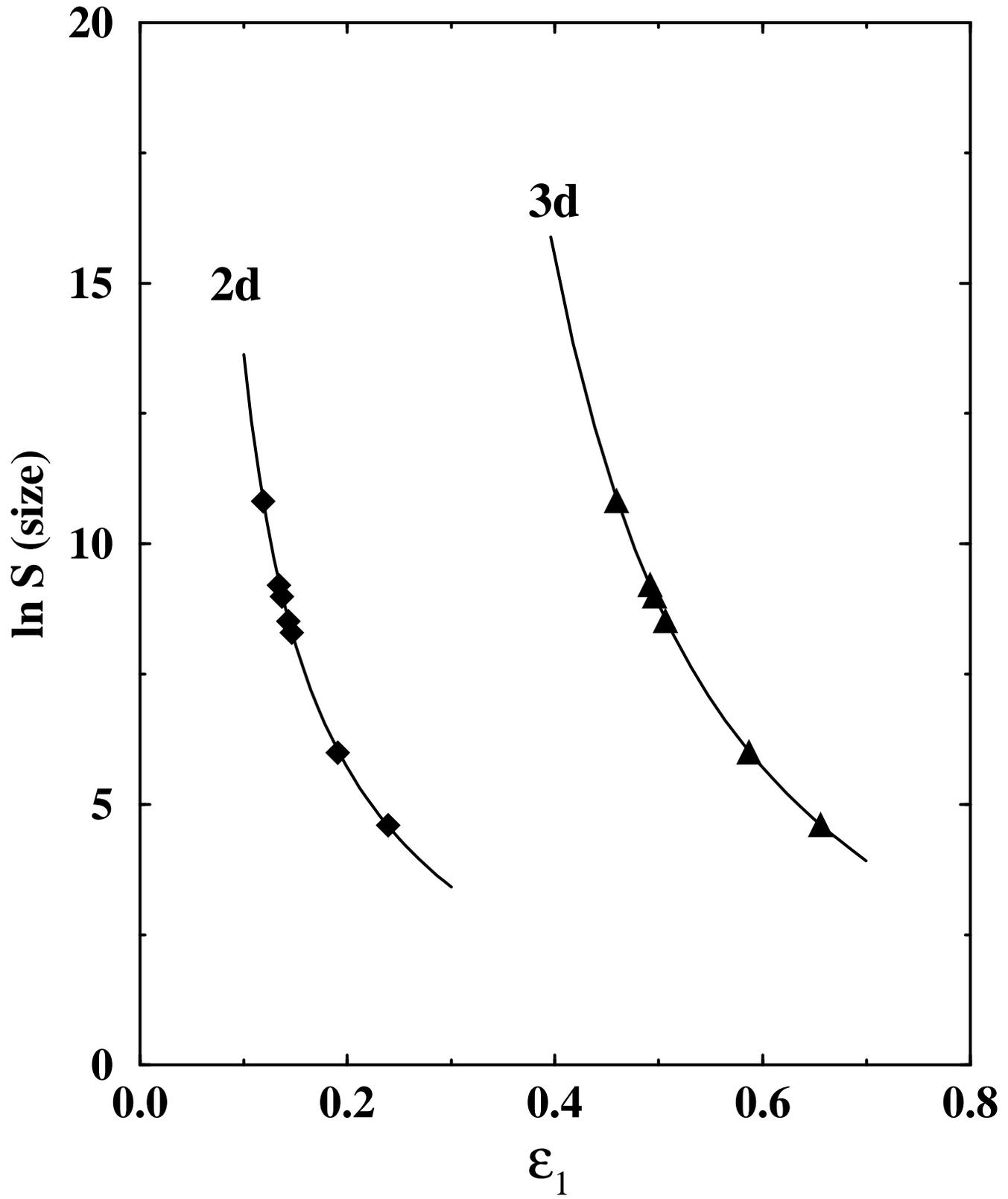
\end{document}